\documentclass{article}
\usepackage[cp1251]{inputenc}
\usepackage{caption}
\usepackage[pdftex]{graphics}
\usepackage{hyperref}
\usepackage{cleveref}
\title{Inflation model and Riemann tensor on non-associative algebra}
\date{}
\author{V. Yu. Dorofeev\thanks{e-mail: friedlab@mail.ru}\\
	A. A. Friedmann Laboratory\\ for Theoretical Physics}
\begin{document}
	\maketitle
\begin{abstract}
In this article the reduction of a $n$-dimensional space to a $k$-dimensional space is considered as a reduction of $N^n$ states to $N^k$ states, where $N$ stands for the number of single-particle states per unit of spatial length. It turns out, this space reduction could be understood as another definition of inflation. It is shown that the introduction of the non-associativity of the algebra of physical fields in a homogeneous space leads to a nonlinear equation, the solutions of which can be considered as two-stage inflation. Using the example of reduction $T\times R^7$ to $T\times R^3$, it is shown that there is a continuous cross-linking of the Friedmann and inflationary stages of algebraic inflation at times $10^{-15}$ with the number of baryons $10^{80}$ in the Universe. In this paper, we construct a new gravitational constant based on a nonassociative octonion algebra.
\end{abstract}


\section*{Introduction}
One of the problems of General relativity is the problem of isotropy of the space around us. Indeed, according to the well-established Friedmann model of the Universe, the indefinite metric of our space-time is defined by a time dependency in the form of
\begin{equation}\label{fv1}
	ds^2 = c^2dt^2 - a^2(t)(dx_1^2 + dx_2^2 + dx_3^2),
\end{equation}
where $a(t)$ is actually the size of our Universe, which is clearly seen in the example of a closed Universe
\begin{equation}\label{vf2}
	ds^2 = c^2dt^2 - a^2(t)(d\xi^2 + sin^2\xi(\sin^2\theta d\varphi^2 + d\theta^2))
\end{equation}
	
According to Friedmann theory until the time when matter became trans\-parent, i. e. since the emergence of the Universe until  $T_{dust}=380$ thousand years or $1.2\cdot10^{13}$ seconds, the equation of Universe's state  looked like $p=\varepsilon/3$ (using historically preceeding approach, which does not include, for example, de Sitter phase with equation of state $p=-\varepsilon$). Thus, in the case of a flat topology of the Universe, the equation for the evolution of the size of the Universe was $a(t)=a_0\sqrt t$. At the moment $T_{dust}$ the equation of state of the Universe changed to $p=0$ and the new equation of the size of the Universe became $a(t)=a_1\root3\of t^2$  or
\begin{equation}\label{vf3}
a(t)=\left\{
	\begin{array}{l}
		a_0\sqrt t,\quad T_{born}\leq t\leq T_{dust.},\\
	  a_1\root3\of {t^2},\quad    T_{dust}\leq t\leq T_{now}			
	\end{array}\right.
\end{equation}
	
The values of the constants $a_0$ and $a_1$ are found from the agreement of modern ideas about the age of the Universe, its current size and continuity of the function $a(t)$. (Again, the power-law nature of the dependency on the time of the radius of the Universe implies the absence of an event horizon, so we consider the visible part of the Universe to coincide with its size. For the de Sitter stage, the horizon arises and is determined by the inverse value of the Hubble constant and, correspondingly, the size of the Universe is much larger than its visible part.) 

From recent observations on PLANK, the modern age of the Universe is estimated at 13.75 billion years or $T_{now}=4.3\cdot10^{17}$ seconds and accordingly the size of the Universe is equal to  $a(T_{now})=1.3\cdot10^{28}$ cm, where
\begin{equation}\label{vf4}
	a_1 = a(T_{now})\cdot T_{now}^{-2/3}=1.3\cdot10^{28}/(4.3\cdot10^{17})^{2/3}=2.3\cdot10^{16}.
\end{equation}
comes from.

Using the continuity $a(t)$ at the moment of changing the equation of state $a_0\cdot T_{dust}^{1/2}=a_1\cdot T_{dust}^{2/3}$, we find $a_0$:
\begin{equation}\label{vf5}
	a_0 = a_1\root6\of {T_{dust}}=2.3\cdot 10^{16}\cdot\root6\of{1.2\cdot10^{13}}= 3.4\cdot 10^{18}.
\end{equation}

As a result we come to the equation of evolution of the radius of the Universe in cm
\begin{equation}\label{vf6}
	a(t)=\left\{
		\begin{array}{l}
			3.4\cdot10^{18}\cdot\sqrt t,\quad T_{born}\leq t\leq T_{dust.},\\ 
			2.3\cdot10^{16}\cdot\root3\of t^2,\quad T_{dust}\leq t\leq T_{now}
		\end{array}	\right.
\end{equation}

Substituting Planck time $10^{-43}$ s as $T_{born}$, we get the initial size of the Universe is $10^{-3}$ cm, and considering that the Planck length is $c\cdot T_{begin}=3\cdot10^{-33}$ cm, we find that at Planck times the universe <<consisted>> of $3.6\cdot10^{29}$ independent pieces. At the same time, according to the latest data, the anisotropy of the Universe appears only in the fourth sign of the accuracy of the temperature of the Universe, depending on the direction.

The most widely accepted way to solve the problem of isotropy of the universe is to quickly inflate. It at times $10^{-35}$ s using the scalar field existing in the early Universe in models of inflation. The history of this approach begins with the works of Gus and is reflected in book \cite{Linde}.
		
The reasons why such a scalar field existed at the beginning of the evolution of the Universe are beyond the scope of these models.

Nevertheless, the inflationary scenario of the early Universe today has a fairly good theoretical basis, consistent with models of strong and weak interactions, and any revision of this approach should not contradict the consequences consis\-tent with the modern theory of elementary particles.
		
Thus, the main results of the Weinberg-Salam theory are related to moments happenden after than $10^{-12}$ s, so perhaps there is some freedom in creating models of the early Universe before these times, although some questions arise due to the beginning of the hadron era.
	
Modern inflationary scenarios are based on model approaches to the evolution of the early Universe. In them, geometry is described by GRG methods, and the matter that induces a geometric structure is described as a scalar field. The main inflationary scenarios in various models are obtained as self-consistent field equations and geometries and can be divided into models of the first and second kind. 
	
1. {\it The model of chaotic inflation}. A primary inflaton field $\varphi(t)$ \cite{Linde2} is postulat\-ed, for which the Euler-Lagrange equation is obtained in curved coordinates as:
\begin{equation}\label{vf7}
	\varphi''+ 3H\varphi'+\frac{dV}{d\varphi}=0
\end{equation}
where $H$ --  Hubble constant. The inflaton field $\varphi$ with potential $V(\varphi)$ appeared in work \cite{Guth} as external to geometry and should have existed at the beginning of evolution.
	
2.  {\it Starobinsky Model}, which takes the Lagrangian $f(R)$ \cite{Starobinsky} with an additional term $R^2$ in addition to the renormalization of the gravitational constant, allowed us to introduce a primary inflaton field, which is already an internal field with respect to geometry.
		
Inflationary scenarios in type I and II models arise from the de Sitter vacuum solution $p= - \varepsilon$ of the Friedman Universe
\begin{equation}\label{vf8}
 \rho'a^3+3(p+\varepsilon)a^2a'=0.
\end{equation}
and <<good>> initial conditions.
		
In this work, an inflation scenario is proposed, which is based on the principle of compacification of the dimensions of the original space. The proposal of this principle is meant to apply it to a non-associative octonion algebra, which could produce insightful results.

\section{Space state}
Efforts to quantize space time date back to works of Bronstein. However, it was most successfully done by Wheeler and de Witt in their joint publication in 1967. In this work, the authors proposed a model for quantization of the Universe based on quantization of the space-time metric. It was shown for the first time that taking into account the nonlinear structure of the Einstein equations, on the basis of which the Hamiltonian of the quantum Universe was constructed, resulted in a loss of time. The Wheeler de Witt equation has the structure of a quantum operator over all possible spatial Riemannian metrics in the representation of the continuum integral \cite{dew}
\begin{equation}\label{f11}
	\left(G_{abcd}\left[\frac\delta{\delta g_{ab}}\right]\left[\frac\delta{\delta g_{cd}}
	\right]-\sqrt g {^{(3)}\!R}\right)\Psi(g_{ab})=0
\end{equation}

Function of the state space of the Universe $\Psi(g_{ab})$ contains an explicit reference to the metric of this space, thereby indicating the character of the early universe as a three-dimensional spatial manifold. In addition, it is obviously possible to proceed from the original space with a large number of dimensions and the subsequent compactification of a part of the dimensions. In any case, the Wheeler de Witt equation points to the spatial character of quantum states.

There is another important quantum approach to the Early Universe the method of multiple interpretation of worlds in the Everett model \cite{ev}. In essence, this model goes back to Fock quantization, when the states of the Universe are interpreted as eigenstates $\xi^{S}_i$ of operators $\hat a,\hat a^+$, defining the general state of the quantum system \cite{grib1}
\begin{equation}\label{f12}
\Psi^S=\sum_{i,j}a_{ij}\xi^{S_1}_i\xi^{S_2}_j
\end{equation}

In contrast to the approach of Wheeler and de Witt, there is initially no dimension of space-time, and the states of the Early Universe are abstract, endowed with an additional space-time structure. On the other hand, in the Fock approach states are characterized solely by a set of quantum variables. Take, for example, a set of various energy states and you will get the quantum energy operator in full accordance with the Bohr interpretation of the atom. 

In this sense, the Aspecs trial is illustrative \cite{aspec}, which shows the quantum state is a state in a certain state space. The peculiarity of such a space of states is that, being spatially unrelated, they nevertheless form a single state that feels changes in each of its parts even when it is separated in coordinate space. This fact has now been experimentally confirmed and has already been applied to the model of quantum signal shifting. 

Due to existence of this kind of analogy, we can further apply Everett's approach to the Early Universe as a space of Fock states. That would allow us to consider the Early Universe in the spirit of the original $n$-dimensional space. This space in the Early Universe can only be the space of Bose particles as fields. This is indicated by the Weinberg-Salam model, according to which Fermi particles are formed due to interaction with the Higgs fields \cite{land1} in its minimum potential energy. Since in the Early Universe all energy is localized in a small physical space, there can be no minimums of energy. However, this is also indicated by the Landau model of a quantum liquid, which is where the analogy with the Higgs fields came from. In the Landau model, the bound states of a quantum liquid are destroyed when the temperature increases. Current temperature does not allow the formation of superconductors at normal temperature \cite{land2}.

The proposed approach to the Early Universe, according to which different dimensions of space-time are possible, turns out to be fruitful in another important way. Thus we can view the reduction of spatial dimensions as an inflation of space-time. This idea emerges if you try to imagine cubes of the same linear size along one of the axes, but with a different number of cells that have the same length for each of their dimensions. In this case, a decrease in the dimension of the space immediately leads to an increase in the effective length at which $n$-dimensional cubes are placed for the same number of $k$-dimensional cubes as cells of possible spatial states of Bose particles in a new space with a smaller number of dimensions. 

It is assumed that the velocity at each coordinate in $n$-dimensional space is stored, but new spatial states are formed in $k$-dimensional space. Fig. 1 illustrates the change in the effective length of the space on the example of the transition from a two-dimensional space to a one-dimensional one. In this case, both the $OX$ axis and the $OY$ axis propagate a wave with the same speed $c$. It is noteworthy that Fig. 1 describes the process of inflation. During this period, no new states are formed -- so the number of states is constant.

\section{Inflation as compactification}
Let's assume that at time $T_0$ there is one state in the space $R^n$ that occupies the volume $V=L_0^n$, $L_0=cT_0=1\cdot T_0$. And allowt for each moment of time $T_0$ for each dimensions of space $R^n$, the number of states increase by one, that is, at time $2T_0$, the number of states is equal to $2^n$, and the volume $V(t)$ occupied by these states at time $t=2T_0$ is equal to $V(t=2T_0)=(2L_0)^n=2^nL_0^n$, and so on.

Let there be the maximum possible number of states in the space $R^n$ per one dimension $R$, equal to $N_0$. Then the maximum possible number of states in the rn space is $N_0^n$. Let's look at the moment when the space $R^n$ is filled with all possible valid states, the states are reduced to the space $R^k\subset R^n(k<n)$.

We define the principle of reduction from the state of the space $R^n$ to the space $R^k$. Since $R^k\subset R^n(k<n)$, the original principle of evolution-the emergence of new states must remain the same, that is, let one state appear in the one-dimensional space of space $R^n$ for a unit of time $T_0$. 

Then $(N_0+1)^n$ states can appear in an $n$ dimensional space per unit of time. Accordingly, the additive is equal to $(N_0+1)^n-N_0^n$. If $N_0>>1$ then $\Delta N_k\sim n N_0^{n-1}$. 

Let this number of states of an $n$-dimensional space pass into states of a $k$-dimensional space. 

At the moment of time, $N_0T_0$ in $k$-dimensional space became $n_0^k=N_0^k$ states. In the next moment, $N_0T_0+T_0$ of them will be $n_1^k=N_0^k+ \Delta N_k\sim N_0^k+n N_0^{n-1}$. Thus, if during a single period $T_0$ in the space $R^n$ the number of states per dimension increased by one, then in the space $R^k$ the number of states will increase by 
\begin{equation}\label{f21}
\Delta n_1\sim\root k\of{N_0^k+nN_0^{(n-1)}}-N_0=N_0\left(\root k\of{N_0^k+nN_0^{n-1-k}}-1)\right).
\end{equation} 

Let $n>k+1$, then 
\begin{equation}\label{f22}
\Delta n_1\sim N_0\left(\root k\of{nN_0^{n-1-k}}\right).
\end{equation}

Thus, the rate of increase of the one-dimensional space in $R^k$ is equal to (\ref{f22}), which can be called the rate of reduction of the space $R^n$ in $R^k$. 

If $N_0$ is large, i. e. $N_0>>n$, then $\Delta n_1^k<<N_0^n$ (remember that $n>k+1$), so all the states of the $k$-dimensional space will not be filled in one period. Since $k$-dimensional space is a subspace of $n$-dimensional space, the number of states in $n$-dimensional space is preserved.

Let's find out the number of states that are formed in $k$-dimensional space in the next unit of time, counting the rate of formation of states in $n$-dimensional space as the same:
\begin{equation}\label{f23}
n_2^k = N_0^k + nN_0^{n-1}+nN_0^{n-1}.
\end{equation} 

Thus, for each time period, the number of new states in the $R^k$ space remains unchanged. 
\begin{equation}\label{f24}
N_0^n/(nN_0^{n-1})=N_0/n.
\end{equation}

The total number of such steps is equal to
\begin{equation}\label{f25}
N_0^n/(nN_0^{n-1})=N_0/n.
\end{equation}

And the total time to fill the space $R^k$ with all states will be $T_0N_0/n$.

The volume $V$ of the space $R^k$ at the moment of its full filling is equal to $V_{end}=(n_{end}L)^k=L^kN_0^n$, and the linear size of the space is equal to $L_{end}=LN_0^{n/k}$

The situation described in this model can be commented on as in figure 2.

Figure 2 shows the starting situation, when all energy is concentrated in a single volume $L_0=cT_0$, equal to the entire volume of the Universe at the initial time $T_0$. Then, during $T_1$, the universe, expanding in a flat n-dimensional space, occupies the volume $L_1^n$. At this point, the expansion in the entire $n$-dimensional space stops, since the minimum energy value has been reached. After this, the process of inflation begins, as shown in fig. 2 by an arc. Thus fig. 2 shows the one-stage process of evolution of the Universe.

\section{A two-stage model of inflation}
Consider the reduction of the space $R^{n+1}$ at $n=7$ to the space $R^{k+1}$-a physical four-dimensional space, when $k=3$ - as cosmological inflation, and we assume that $R^{n+1}=T\otimes R^n$ and $R^{k+1}=T\otimes R^k$. 

Let's consider the next stages of the evolution of the early Universe.

1. The evolution of the Universe occurs when a single Bose state appears at the moment $T_0=1/(Nk_0)$. Since the total energy of all states is equal to $E_{all}=N^4k_0^2$ (\ref{f24}), and the energy of one state is equal to $Nk_0$, then $N_0^7=N^3k_0$ and $L_0=c\cdot T_0$.

2. Inflation occurs when all single states of the Universe are constructed in the space $R^n$: $L_{1b}=c\cdot T_{EUIbegin}=c\cdot N_0T_0=c\cdot N^{3/7}T_0$. 

3. The size of the Universe before the first inflation is $L_{1b}=c\cdot T_{EUIbegin}=c\cdot N_0T_0=c\cdot N^{3/7}T_0$.

4. During inflation, for equal periods of time, the physical size of the Universe increases by the same amount, so there is a linear law of growth of the size of the Universe.

5. The duration of the first inflation is equal to $T_{EUItime}=N_0T_0/n=N^{3/7}T_0/n$.

6. The first stage of inflation ends at time $T_{EUIend}=(n+1)N^{3/7}T_0/n$.

7. The Linear size of the Universe at the end of the first inflationary period of evolution is $L_1=L_{EUI}=L_0N_0^{7/3}=L_0N$.

8. The Number of states of particles in the Universe at the end of the first period of inflation becomes equal to $N^3$, and in each state there is one particle.

After inflation, a new situation arises. The space size has increased and a new solution (\ref{f15}) appears with $N=1$ in (\ref{f24}). The total number of states after the first stage of evolution is $N_1^7=N^3$. Since the particles are free, this space is again $M_8$, and the number of states becomes $2^n$ now after $T_1=N_1T_0=N^{3/7}T_0$.

9. Repeating the previous arguments, we get that in the space $R^7$ all the states of $N_1^7$ will be filled at the moment $T_{EUIIbegin}=N_1\cdot T_1=N^{3/7}T_1$ and the second non-inflationary period of the evolution of the Universe will be over.

10. By the end of the second non-inflationary period, the size of the Universe will increase to $L_{EUII}=c\cdot T_{EUIIbegin}=c\cdot N^{3/7}T_1$.

11. At the time $T_{EUIIbegin}$, the second inflationary period, which will continue in $T_{EUIItime}=N_1T_1/n=N^{3/7}T_1/n$. 

12. The linear size of the Universe to the end of the second inflation period will begin of evolution increased to $L_{EUII}=NL_1$.

13. The second period of inflation, will end at $T_{EUIIend}=(n+1)N^{3/7}T_1/n$.

\begin{equation}\label{f31}
a(T_{EUIIend})=L_{EUII}=NL_1=N^2 L_0 =cN^2T_0
\end{equation} 
or
\begin{equation}\label{f32}
\frac87a_0\sqrt{T_0}N^{3/7}=cN^2T_0
\end{equation}
jnrelf
\begin{equation}\label{f33}
N^{4-6/7}=\frac{64}{49c^2}a_0^2T_0^{-1}
\end{equation}
be fulfilled.

\begin{equation}\label{f34}
a(T_{EUIIend})=L_{EUII}=NL_1=N^2 L_0 =cN^2T_0
\end{equation} 
or
\begin{equation}\label{f35}
\frac87a_0\sqrt{T_0}N^{3/7}=cN^2T_0
\end{equation} 
then
\begin{equation}\label{f36}
N^{4-6/7}=\frac{64}{49c^2}a_0^2T_0^{-1}
\end{equation} 

Substituting $T_0=10^{-43}$ s, $c=3\cdot10^{10}$ cm/s and $a_0$ from (\ref{f31}), we get
\begin{equation}\label{f37}
N=0.8\cdot10^{19}
\end{equation} 

For $N=10^{19}$, we find that inflation II ended in $8.2\cdot10^{-14}$ s, and the size of the Universe was $6.5\cdot10^{11}$ cm. 

From 1. - 13. it is not difficult to find that at the initial moment of $T_0=10^{-43}$ s, the linear size of the Universe is $L_0=3\cdot10^{-33}$ cm, and the total number of states with an energy of $E_0=10^{19}k_0$ in each was $N^3=10^{57}$. At the moment of $T_{EUIbegin}=1.4\cdot10^{-35}$ s, when the linear size of the Universe becomes $4.2\cdot10^{-25}$ cm, the first inflation will occur with a duration of $0.2\cdot10^{-35}$ s, and the linear size of the Universe will increase to $3\cdot10^{-14}$ cm. The second inflation will start at $T_{EUIIbegin}=7.2\cdot10^{-14}$ s, last $10^{-14}$ s, and end at $T_{EUIIend}=8.2\cdot10^{-14}$ s, while the size of the Universe will increase from $3\cdot10^{-14}$ cm to $6.5\cdot10^{11}$ cm during the second stage. 

The linear size of the Universe and the beginning of the Friedman stage of the evolution of the Universe were well consistent with the classical model. But there is a serious contradiction to the proposed model with the total number of particles, if we consider the model $N^4=10^{76}$ as the number of baryons, while the energy $E_0=10^{19}k_0$, which follows from the model, agrees very well with the Planck mass. The situation can be corrected if we do not try to link the Friedman and algebraic stages of inflation, immediately putting $N^4=10^{80}$-a number that reflects the current idea of the number of baryons in the Universe. In this case, the algebraic stage of evolution ends in $T_{EUIIend}=3.1\cdot10^{-12}$ s. The size of the Friedman universe at this moment is $4.4\cdot10^{12}$ cm, and the algebraic Universe is $4.4\cdot10^{14}$ cm. Reducing the initial time of $10^{-43}$ to $10^{-46}$ s allows you to stitch the Universe with a size of $1.4\cdot 10^{11}$ cm at the end of the al-Hebraic inflation in $3.1\cdot10^{-15}$ s.

From (\ref{f33}) we get
\begin{equation}\label{f38}
N^{3\frac17}T_0=\frac{64}{49c^2}a_0^2
\end{equation}

\section{Algebra of the Early Universe}
We set the task to find the field equation in the early Universe.

Let the state space of the early Universe be described by some algebra, which we define on some linear space. We define the dual space as a coordinate space. The space dual to the coordinate space is defined as a vector space. If we take the physical space $T\times R^3$ as the coordinate space, we can, for example, use its spinor representation \cite{Penrose}.

$$[physical\, world]\longrightarrow V(\hbox{vector space})$$
$$V\longrightarrow G=U(1)\times SU(2),\quad V\longrightarrow V^*,\quad V^*\longrightarrow V^{**}$$
$\forall v\in V:v=v_0\sigma^0+v_1\sigma^1+v_2\sigma^2+v_3\sigma^3=
\left(\matrix{v_0-v_3&g_1+iv_2\cr v_1-iv_2&v_0+v_3}\right)$

As a result, we get:

$V^*$ -- dual coordinate covector space of Minkowski and
$$\forall x\in V^*:x_\mu=x_\mu^0\sigma^0+x_\mu^1\sigma^1+x_\mu^2\sigma^2+x_\mu^3\sigma^3$$
or
\begin{equation}\label{f41}
x_0^0=cT,x_1^1=X,x_2^2=Y,x_3^3=Z,\quad x=
\left(\matrix{cT-Z&X+iY\cr X-iY&cT+Z}\right)
\end{equation} 

As a result, we get a model that basically coincides with the usual quantum field theory (the implementation of the corresponding Penrose program has a number of important results and is currently developing). 

It is known that a group with a layer - by-layer transition can be realized on a $V^{**}$ - vector space, 
\begin{equation}\label{f42}
A_\mu'=g(x) A_\mu(x)g^{-1}(x)+\partial_\mu g(x) g^{-1}(x)
\end{equation}
where $g(x)$ is a group element $U(1)\times SU(2)$ that matches vector fields on the algebra $u(1)\times su(2)$ at different group points with the same coordinate $x\in R^4$. 

We research the vector field $A_\mu(x)$ on the algebra $u(1)\times su(2)$ by \cite{Izikson}. To do this, consider the isometry of
$$Pe^{\oint_lA_\mu dx^\mu}=I+\oint_lA_\mu dx^\mu+P\int\int_{x_2>x_1} A_\mu(x_2)dx^\mu_2A_\nu(x_1)dx_1^\nu+o(l^2)=$$
\begin{equation}\label{f43}
=I+\oint_{\partial\Omega}A_\mu dx^\mu+\frac12\oint_{\Omega}F_{\mu\nu}dx^\mu\wedge dx^\nu+o(l^2)
\end{equation} 

For further research, we will write out the corresponding coordinate representations of the above equations.
\begin{equation}\label{f44}
F_{\mu\nu}=\partial_\mu A_\nu-\partial_\nu A_\mu-[A_\mu,A_\nu]
\end{equation}

In the case of commutative fields on the group $U(1)$, the tensor (\ref{f44}) has the form of a coordinate function $F_{\mu\nu}^0=\partial_\mu A_\nu^0-\partial_\nu A_\mu^0$. On the group $SU(2)$, the tensor $F_{\mu\nu}$ is written in layers: $F_{\mu\nu}^a=\partial_\mu A_\nu^a-\partial_\nu A_\mu^a-([A_\mu,A_\nu])^a$. 

The covariant derivative $\nabla$ is formed by raising the index $\mu$ using the metric tensor $g^{\mu\nu}$. Metric tensor $g^{\mu\nu}$ appears as a transition to arbitrary coordinates in the original flat Minkowski space $R^4$ on layer $a$, so the index of layer $a$ remains upper, and the coordinate index $\mu$ rises from the bottom to the top. As a result
\begin{equation}\label{ninf18}
\nabla^\mu F_{\mu\nu}=\partial^\mu F_{\mu\nu}-[A^\mu,F_{\mu\nu}]=0
\end{equation}
or
\begin{equation}\label{f45}
\partial^\mu\partial_\mu A_\nu-\partial^\mu\partial_\nu A_\mu-[A^\mu,\partial_\mu A_\nu-\partial_\nu A_\mu- [A_\mu,A_\nu]]- \partial^\mu[A_\mu,A_\nu]=0
\end{equation}

As follows from \cite{Izikson}, only the physical part of
\begin{equation}\label{f46}
\oint A_\mu dx^\mu=\oint(A_\mu^{||}+A_\mu^{\bot})dx^\mu=\oint A_\mu^{||}dx^\mu
\end{equation}
can be left due to transformations (\ref{f42}). 

Therefore, we consider 
\begin{equation}\label{f47}
		A_\mu^a(x)=A_\mu^{||}(x)=n_\mu^a A(x),\qquad x=k_0x^0-\vec k\vec x
\end{equation}
with a normalized $(n_\mu^a)^2=1$ vector in $M^4$ space. As a result, we come to the equation for transverse vector fields $k_\mu A^\mu=0$
\begin{equation}\label{f48}
			\partial^\mu\partial_\mu A(x)=0
\end{equation}

Note that in (\ref{f48}), in accordance with \cite{Izikson}, the Greek index $\mu$ numbers the real numbers of the $dx^\mu$ coordinates in $R^4$. the Group index carries a field, for example, in the case of the group $SU(2)$, we get $F_{\mu\nu}=F_{\mu\nu}^a\sigma^a$. In our case, this is not the case for the simple reason that the coordinate space is constructed as dual to the vector space, so it turns out to be from the same algebra. In a sense, this is not new-Penrose's twistor approach just means an attempt to construct a physics in which the coordinate space is an element of the $SU(2)$ algebra. 

When finding the tensor $F_{\mu\nu}$ in (\ref{f46}), the commutativity and associativity of the field $A_\mu$ and the coordinates $dx^\mu$ are used, so they must be output in a new way.

Mapping a space element with a basis $e$ leads to a component record of the vector $A=A_\mu e^\mu=A_\mu^a\Sigma^ae^\mu$ and, respectively, the covector $x=x^\mu e_\mu=x^{\mu(b)}e_\mu\Sigma^b$.

Therefore
\begin{equation}\label{f49} 
A\cdot dx=A_\mu dx^\mu=A_\mu^adx^{\mu(b)}\Sigma^a\Sigma^b
\end{equation}

Note that in (\ref{f49}) $A_\mu^ax^{\mu(b)}$ are real numbers, and $\Sigma^a\Sigma^b$ are the basis elements of the algebra. 

When passing from point $x$ to point $x'=x+\Delta x$, the vector $V$ changes due to the change of the coordinate and the element of the algebra
\begin{equation}\label{f410} 
\delta V(x)=V_{,\mu}\Delta x^\mu-VA_\mu(x)\Delta x^\mu
\end{equation}

Therefore, as shown in \cite{Izikson}, we could come to  (\ref{f43}). However, it is necessary to take into account the non-associativity of the algebra. To do this, we calculate $\oint A_\mu dx^\mu$ to the second-order infinitesimal accuracy using (\ref{f49}) and taking the vector components of $A_\mu^a\Sigma^a$ as a vector $V$ in (\ref{f410}). 
$$\oint A_\mu dx^\mu=A_\mu(x)\Delta x^\mu_1+A_\mu'(x+\Delta x_1)\Delta x^\mu_2-A_\mu''(x+\Delta x_2)\Delta x^\mu_1-A_\mu(x)\Delta x^\mu_2=$$
$$+(A_\mu(x)-(A_\mu(x+\Delta x_2)-A_\mu(x)A_\nu(x)\Delta x^\nu_2))\Delta x^\mu_1-$$
$$-(A_\mu(x)-(A_\mu(x+\Delta x_1)-A_\mu(x)A_\nu(x)\Delta x^\nu_1))\Delta x^\mu_2=$$
$$=((\partial_\nu A_\mu+A_\mu A_\nu)\Delta x^\nu_1))\Delta x^\mu_2-
((\partial_\nu A_\mu+A_\mu A_\nu)\Delta x^\nu_2))\Delta x^\mu_1=$$
$$=(\partial_\nu A_\mu+A_\mu(x)A_\nu))
(\Delta x^\nu_1\Delta x^\mu_2-\Delta x^\nu_2\Delta x^\mu_1)+$$
$$+(((\partial_\nu A_\mu+A_\mu A_\nu)\Delta x^\nu_1))\Delta x^\mu_2-
(\partial_\nu A_\mu+A_\mu A_\nu)(\Delta x^\nu_1\Delta x^\mu_2))-$$
$$-(((\partial_\nu A_\mu+A_\mu A_\nu)\Delta x^\nu_2))\Delta x^\mu_1-
(\partial_\nu A_\mu+A_\mu A_\nu(\Delta x^\nu_2\Delta x^\mu_1)=$$
\begin{equation}\label{f411} 
=\frac12(\partial_\nu A_\mu-\partial_\mu A_\nu-[A_\nu,A_\mu])
(\Delta x^\nu_1\Delta x^\mu_2-\Delta x^\nu_2\Delta x^\mu_1)+NA
\end{equation}
\begin{equation}\label{f412} 
NA=(((\partial_\nu A_\mu+A_\mu A_\nu)\Delta x^\nu_1)\Delta x^\mu_2-
(\partial_\nu A_\mu+A_\mu A_\nu)(\Delta x^\nu_1\Delta x^\mu_2))-$$
$$-(((\partial_\nu A_\mu+A_\mu A_\nu)\Delta x^\nu_2)\Delta x^\mu_1-
(\partial_\nu A_\mu+A_\mu A_\nu)(\Delta x^\nu_2\Delta x^\mu_1))
\end{equation}

The first term in (\ref{f412}) corresponds to the usual stress tensor and has the form $F_{\mu\nu}(dx^\mu\wedge dx^\nu)$ of a non-Abelian associative field. Note that the parentheses indicate the order of operations on the one hand and the two dual structures on the other hand. The term $NA$ makes additional corrections due to the non-associativity of the vector and covector spaces. We separate the dual spatial part from the vector fields in the term $NA$ by writing the expression in the base.
$$NA=\frac12[(((\partial_\nu A_\mu^a-\partial_\mu A_\nu^a)(\Delta x^{\nu(b)}_1\Delta x^{\mu(c)}_2-\Delta x^{\nu(b)}_2\Delta x^{\mu(c)}_1)\cdot$$
$$((\Sigma^a\Sigma^b)\Sigma^c-\Sigma^a(\Sigma^b\Sigma^c))+
(A_\mu^aA_\nu^b-A_\nu^aA_\mu^b)(\Delta x^{(c)\nu}_1\Delta x^{(d)\mu}_2-\Delta x^{(c)\nu}_2\Delta x^{(d)\mu}_1)\cdot$$
$$(((\Sigma^a\Sigma^b)\Sigma^c)\Sigma^d-(\Sigma^a\Sigma^b)(\Sigma^c\Sigma^d))]=$$
$$=(\varepsilon^{adck}(\partial_\nu A_\mu^a-\partial_\mu A_\nu^a)+\varepsilon^{abl}
\varepsilon^{lcdk}(A_\mu^kA_\nu^c-A_\nu^kA_\mu^c))\cdot$$
$$\cdot(\Delta x^{(c)\nu}_1\Delta x^{(d)\mu}_2-\Delta x^{(c)\nu}_2\Delta x^{(d)\mu}_1)\Sigma^k$$ 
where $\varepsilon^{abcd}$ -- associator of basic elements:
$$(\Sigma^a\Sigma^b)\Sigma^c-\Sigma^a(\Sigma^b\Sigma^c)=2\varepsilon^{abcd}\Sigma^d$$
in which there is no summation by repeating indexes.

\section{Construction and research of field equations on nonassociative algebra}
We will look for a solution in the space $T\otimes R^n$ with metric
\begin{equation}\label{f51}
		ds^2=c^2dt^2 - dx_1^2 -\dots-dx_n^2=dx_0^2-dx_1^2 -\dots-dx_n^2,
\end{equation}
for the vector field $A_\mu(x)$ of the algebra $su(n+1)$ in the form \cite{Fadeev}
\begin{equation}\label{f52}
		\nabla^\mu F_{\mu\nu}=\partial^\mu F_{\mu\nu}-[A^\mu,F_{\mu\nu}]=0
\end{equation}	
or		
		\begin{equation}\label{f53}
		\partial^\mu\partial_\mu A_\nu-\partial^\mu\partial_\nu A_\mu-[A^\mu,\partial_\mu A_\nu-\partial_\nu A_\mu- [A_\mu,A_\nu]- \partial^\mu[A_\mu,A_\nu]=0
		\end{equation}

We will search for wave homogeneous transverse solutions of the form
$A_\mu^a=n^a_\mu A(x),n^a_\mu n^b_\mu=\delta^{ab}, n_\mu^ak^\mu=0, x=k_0x^0-\vec k\vec x$.

In the case of associative algebra, there is only a homogeneous solution of the form $\partial^\mu\partial_\mu A_\nu=0$.

But for nonassociative fields, a new term appears and the equation takes the form
\begin{equation}\label{f54}
			\partial^\mu\partial_\mu A(x)+A^3(x)=0
\end{equation}

Let the conditions
\begin{equation}\label{f55}
			A''(x_0)+A^3(x_0)=0
\end{equation} 
and
\begin{equation}\label{f56}
			A(x_0=0)=1,\quad A'(x_0=0)=0
\end{equation}
or
		\begin{equation}\label{f57}
		A(x_0=0)=0,\quad A'(x_0=0)=1/\sqrt2.
		\end{equation} 
be set at the point $\vec x=0$. 

Equation (\ref{f55}) is converted to the form
\begin{equation}\label{f58}
		\partial^0(\frac12(\partial_0A(x_0))^2+\frac14A^4(x_0)-C^4)=0.
\end{equation}

The solution (\ref{f58}) for $C=1/\sqrt2$ is an elliptic Jacobi function $sn(x_0|1/\sqrt2)$, which for initial conditions (\ref{f55} -- \ref{f56}) are conveniently is denoted as $cots(x_0)$ and $sitn(x_0)$, respectively. Both functions $cots(x_0)$ and $sitn(x_0)$ are periodic with a period $T_{sn}=\frac{4\sqrt{2\pi}\Gamma(5/4)}{\Gamma(3/4)}\sim6.4163>2\pi$.

In Fig. 4, the functions $cots(x_0)$ and $sitn(x_0)$ are combined with the trigonometric functions $\cos(k_0x_0)$ and $\sin(k_0x_0)$, so it is almost certain that $cots(x_0)=\cos(k_0x_0)$ and $sitn(x_0)=\sin(k_0x_0)$ where $k_0=2\pi/T_{sn}\sim0.847$.

(\ref{f54}) we can compare the Lagrangian density $L$ of a scalar field a of the form $A(x)$
\begin{equation}\label{f59}
		L=\frac12\partial^\mu A(x)\partial_\mu A(x)-\frac14A^4(x)
\end{equation}
which corresponds to the Hamiltonian density $H=(\partial_0A)^2-L$ as
\begin{equation}\label{f510}
		H=\frac12\partial^0A(x)\partial_0A(x)+\frac12(\nabla A(x))^4+\frac14A^4.
\end{equation}

As noted above, solutions (\ref{f55} -- \ref{f57}) are close to trigonometric functions $A_{tr}(x)$ of the form $\cos(k_0x_0)$ and $\sin(k_0x_0)$, so replace (\ref{f59}) with
\begin{equation}\label{f511}
		L=\frac12\partial^\mu A_{tr}(x)\partial_\mu A_{tr}(x)-\frac{m^2}2 A_{tr}^2(x)
\end{equation}
and
\begin{equation}\label{f512}
		H=\frac12\partial^0A_{tr}(x)\partial_0A_{tr}(x)+\frac12(\nabla A_{tr}(x))^2+\frac{m^2}2 A_{tr}^2(x).
\end{equation}
the constant $C^4$ in (\ref{f512}) based on (\ref{f58}) can be interpreted as the energy density of the scalar field.

The Lagrangian (\ref{f511}) can be quantized by assuming $A_{tr}$ as an operator
\begin{equation}\label{f513}
		\hat A_{tr}(x)=\sum_ke^{ikx}\hat a(k)+e^{-ikx}\hat a^+(k)
\end{equation}
for which the quantization condition is met in the pulse representation
\begin{equation}\label{f514}
		[a(k),a^+(k')]=\delta_{kk'}.
\end{equation}

For $C=N/\sqrt2$, the solution (\ref{f55}) is the function 
\begin{equation}\label{f515}
		A(x_0)=N sitn(Nx_0)
\end{equation}
that is, the solution is a single mode, which, in accordance with Fig. 4, can be compared with single-mode harmonic oscillations with the total energy
\begin{equation}\label{f516}
		E=N^4k_0^2.
\end{equation}	
		
This is a solution to the main mode of the form $A(x_0)$.

Since the first instant of the Universe corresponds to the minimum time, which cannot be less than the time of natural vibrations of a physical particle and is equal to $T_0=(Nk_0)^{-1}$, we have obtained that at the very first moment the universe is concentrated in one state. This state contains $N^3k_0\sim N^3 $ particles. The energy of each such particle is equal to $Nk_0$. Thus, the total energy of the Universe in the first instant of the Universe is equal to (\ref{f24}).

On the other hand, the original equation is the wave equation, so the solution is the wave. Therefore, the physical linear size of the Universe at the first moment is equal to $L_0=cT_0$.

In the next few moments, the size of the Universe will increase. However, by virtue of the law of conservation (\ref{f58}), the solution (\ref{f515}) is preserved. The size of the Universe will increase, so the density of the number of particles per unit length in the space $R^n$ will fall to a certain number $N_0$ -- when there is exactly one particle in the space $R^7$in the same state. Previously, this linear space size $R^7$ was calculated as $L_1$, and the corresponding evolutionary period was called the first stage.

The first inflation phase is a transition from the solution of (\ref{f515}) the first evolutionary phase the second evolutionary phase
\begin{equation}\label{f517}
		A(x_0)=sitn(x_0)
		\end{equation}
which at the point $x=0$ is mapped to harmonic new massive scalar particle, by analogy with the first step
		\begin{equation}\label{f518}
		A_{tr}(x_0)=a_k\sin(k_0x_0),\quad m=k_0
		\end{equation}

Note that now the energy is not $Nk_0$, but just $k_0$.

If the first scalar particle is understood as the Planck Higgs ($m=Nk_0$), then the second scalar particle is an ordinary Higgs boson with ($m=k_0$) the same evolutionary logic of the Early Universe.

We can assume that by analogy with the Higgs boson, Planck bosons of Higgs particles form Planck electrons, Planck protons and Planck neutrons. In the future, Planck particles can form Planck atoms, which are a good candidate for the dark matter of the Universe. In this regard, it is interesting to study the possibility of the formation of Planck stars in our Universe and their further evolution.

The second evolutionary stage of the Universe ends at the moment when there is only one particle with mass $k_0$ in each state of space $R^7$. At this point, the second inflationary stage begins until there is exactly one particle in each state of the space $R^3$. In the future, the four-dimensional space-time manifold evolves, and the metric becomes Riemannian.

\section{The construction of the Riemann tensor}
Back to (\ref{f411}). Let's write a nonassociative quadratic part over a vector field.
\begin{equation}\label{f61}
\varepsilon^{abl}\varepsilon^{lcdk}(A_\mu^aA_\nu^b-A_\nu^aA_\mu^b)\cdot
(\Delta x^{(c)\nu}_1\Delta x^{(d)\mu}_2-\Delta x^{(c)\nu}_2\Delta x^{(d)\mu}_1)\Sigma^k\end{equation}

By construction, we consider the nonassociative part corresponding to the Calley octave algebra, so we write non-zero values of the fully antisymmetric associator $\varepsilon^{lcdk}$ and also the fully antisymmetric structural constants $\varepsilon^{abc}$ for octonions
\begin{equation}\label{f62}
\varepsilon^{123}=\varepsilon^{145}=\varepsilon^{176}=\varepsilon^{246}=
\varepsilon^{257}=\varepsilon^{347}=\varepsilon^{365}=1,
\end{equation}
\begin{equation}\label{f63}
\varepsilon^{1247}\;=\;\varepsilon^{1265}\;=\;\varepsilon^{2345}
\;=\;\varepsilon^{2376}\;=\;\varepsilon^{3146}\;=\;\varepsilon^{3157}
\;=\;\varepsilon^{4567}\;=1
\end{equation}

It follows from (\ref{f62}--\ref{f63}) that the associator is different from zero only if at least two of its elements belong to the algebra of the higher fields (the higher fields have indexes from 4 to 7). If all indices belong to older fields, and the indexes of the structural constants in (\ref{f62}) must belong to the younger fields (lower field have an index from 1 to 3), which should not be, as younger fields must be vector representations of the boson algebra $su(2)$. Hence the index of $l$ in (\ref{f61}) must belong to the younger field, but then the index $k$ must be for the younger fields

It was previously shown that there is a homogeneous solution for higher fields on a nonassociative algebra in the form of Jacobi functions (\ref{f54}), so we can assume that
\begin{equation}\label{f66}
\Delta x^{(c)\nu}_i=sitn(x)\Delta\tilde x^{(c)\nu}_i,\qquad i=1,2
\end{equation}
where $\Delta\tilde x^{(c)\nu}_i$ it is already a constant value, and $sitn(x)$ is a Jacobi function. Enter a new value
\begin{equation}\label{f67}
R^{\kappa\lambda}_{\mu\nu}(k)\Sigma^k=\varepsilon^{(\kappa+4)(\lambda+4)l}\varepsilon^{l(\nu+4)(\mu+4)k}(A_\mu^{\kappa+4}A_\nu^{\lambda+4}-A_\nu^{\kappa+4}A_\mu^{\lambda+4})\Sigma^k
\end{equation}

Since the isotropic solution is considered, the introduced tensor is the same for all indices $k$ and is equal to just $R^{\kappa\lambda}_{\mu\nu}=R^{\kappa\lambda}_{\mu\nu}(k)$.

From the form (\ref{f67}), we get the properties $R^{\kappa\lambda}_{\mu\nu}$:
\begin{equation}\label{f68}
R^{\kappa\lambda}_{\mu\nu}=-R^{\lambda\kappa}_{\mu\nu}=-R^{\kappa\lambda}_{\nu\mu}
\end{equation}

Omitting indexes using the metric tensor $g_{\mu\nu}$ of some space-time $R^4$, from the duality of vector and coordinate variables for an isotropic solution, we can additionally assume that
\begin{equation}\label{f69}
R_{\kappa\lambda\mu\nu}=R_{\mu\nu\kappa\lambda}
\end{equation}

Thus, a tensor with the properties of the Riemann tensor appears on a nonassociative algebra. As you can see, the second term in (\ref{f412}), which contains derivatives of vector fields, does not have the necessary properties.

Thus, the nonassociative term of the Wigner loop (\ref{f411}) on a nonassociative algebra can be rewritten as
\begin{equation}\label{f610}
\oint A_\mu dx^\mu=sitn^2(x)R^{\kappa\lambda}_{\mu\nu}(k))\Delta S^{\nu\mu}_{12}\Sigma^k+\dots
\end{equation}
where all indexes take values from 0 to 3.

Since the loop rushes to the point, we enter the notation
\begin{equation}\label{f611}
\frac1{\tilde T_{sn}}\int_0^{\tilde T_{sn}}sitn^2(Nx)dx=\frac1{N^2T_{sn}}\int_0^{T_{sn}}sitn^2(x)dx=\frac1{2\kappa}
\end{equation}
where $\kappa$ is the gravitational constant.

Thus, the Riemann tensor is obtained on a four-dimensional space-time ($\mu,\nu,\kappa,\lambda=0,1,2,3$) with the desired gravitational constant, and thus the justification for the Friedman stage appears (see Fig. 3) after $T_{EUIIe}$ early universe.

\section{Conclusion}
The solution for a vector field of the form (\ref{f54}) in an isotropic space was obtained not only by introducing non-associativity of the interaction, but also by using the idea of Penrose \cite{Penrose} to represent coordinates on an external algebra. However, it turned out that associative algebra did not give a nonlinear interaction in a homogeneous space. But the idea of the duality of vector and co-vector fields, according to which the algebra of field representation can also coincide, has become fruitful.

Solution (\ref{f12}) shows that in the early Universe, in addition to free massless wave solutions, there is a solution that can be interpreted as a mass solution. It turns out that two types of solutions are clearly distinguished as solutions with two masses: one solution can be understood as Planck particles and the second solution-hadron particles. Formally, intermediate states cannot be excluded and this issue requires further research. Planck's massive solutions occur at times of $10^{-35}$ seconds. If these solutions are considered as Planck-type Higgs bosons, then formally such solutions can form part of a massive Universe of approximately equal mass to the visible part. This part of the Universe could represent dark matter particles.

\section{Acknowledgements}
I am grateful to RFBR for financial support under grant No. 18-02-00461 ''Rotating black holes as the sources of particles with high energy''.

\end{document}